\documentclass[twocolumn]{aastex631}

\begin{document}

%\title{Detection of faint sources in UVIT using Poisson distribution of background}
\title{Detection of Faint Sources by the UltraViolet Imaging Telescope Onboard AstroSat
Using Poisson Distribution of Background}

\author[0001-0001-4279-7041]{B. Ananthamoorthy}
\affiliation{Manipal Centre for Natural Sciences, Manipal Academy of Higher Education, Manipal, India}

\author[0000-0002-5123-972X]{Debbijoy Bhattacharya}
\affiliation{Manipal Centre for Natural Sciences, Manipal Academy of Higher Education, Manipal, India}

\author[0000-0001-8201-8148]{P. Sreekumar}
\affiliation{Manipal Centre for Natural Sciences, Manipal Academy of Higher Education, Manipal, India}

%\author{C.S. Stalin}
%\affiliation{Indian Institute of Astrophysics, Bangalore, India}

\author{Swathi B}
\affiliation{Manipal Centre for Natural Sciences, Manipal Academy of Higher Education, Manipal, India}
\affiliation{Space Physics Laboratory, Vikram Sarabhai Space Centre, Thiruvananthapuram, Kerala, India}

\correspondingauthor{Debbijoy Bhattacharya}
\email{debbijoy.b@manipal.edu}



\begin{abstract}

We present an improved approach for constructing the UV source catalogs using observations from the UltraViolet Imaging Telescope (UVIT) onboard {\sl AstroSat}, by considering the Poisson distribution of the UV background. The method is tested extensively using fields that are not crowded, the Small Magellanic Cloud (SMC) and M31 (Field 13). The results are compared with previous studies that used UVIT observations. This approach is successful in detecting fainter sources and produces a large number of new sources ($\sim 15$ to $92\%$  more). Most of the newly discovered UV sources fall in the faint end of the source distribution (m $\gtrsim 22$). The counterparts at other wavelengths are identified for most sources. This approach is more efficient for source detection and provides an opportunity to explore new classes of UV sources. 

\end{abstract}


%% and \citet commands to identify citations. The citations are

\section{Introduction} \label{sec:intro}

Ultraviolet (UV) radiation originating from astrophysical sources offers valuable information for understanding the various processes in these objects. It is an important tracer of star formation and the effect of intrinsic dust extinction in galaxies \citep[e.g.,][and references therein]{Calzetti2013, Salim2007, Kong2004, Chander1999, Bianchi1999, Kennicutt2012}. It also plays a pivotal role in exploring our galaxy's stellar population, including massive stars and objects like Planetary Nebulae (PNs), Cataclysmic Variables (CVs), blue straggler stars (BSSs), and white dwarfs (WDs) \citep[e.g.,][]{Bianchi2011, Rao2021, Gomez2023, Subramaniam2020, Sion1999}. The study of active galaxies in UV provided information in characterizing the `big blue bump' \citep{Sanders1989}, likely to be originating from accretion disks in Active Galactic Nuclei (AGN) \citep[e.g.,][]{Shields1978, Ward1987}, and also emission from jets \citep[e.g.,][]{Bhattacharya2021,Gulati2023}. Astrophysical UV background is of great importance, and its origin is still debated \citep[e.g.,][]{Murthy2010, Henry2015}. 

{\it Galaxy Evolution Explorer} ({\it GALEX}) \citep{Martin2005, Morrissey2007}, {\it X-ray Multi Mirror Mission}-Optical Monitor ({\it XMM}-OM) \citep{Mason2001}, Ultra-Violet Optical Telescope onboard {\it Swift} ({\it Swift}-UVOT) \citep{Mason2004, Roming2005}, and Wide Field Camera 3 (WFC3) onboard {\it Hubble Space Telescope} (HST) \citep{2008SPIE.7010E..1EK} are among the major UV instruments in the recent past that have provided insights into the UV sky. UltraViolet Imaging Telescope (UVIT) \citep[e.g.,][]{Kumar2012, Subramanian2016, Tandon2017, Tandon2020} onboard {\it AstroSat}, with an angular resolution of $\sim 1.2$-$1.5$$^{\prime\prime}$ is the only telescope that is currently operational in the FUV range ($120$nm-$180$nm) with nearly $3$-$4$ times better resolution than GALEX (angular resolution of $\sim 4.5$-$5$$^{\prime\prime}$) \citep{Morrissey2007}. 

With an improved angular resolution, it is important to construct a UVIT source catalog extending to a fainter limit. Multiple epoch observations will also provide a platform for the temporal variability study of these sources \citep[e.g.,][]{Leahy2021v}. 

The background calculation is important for the proper detection of sources and their photometry from astronomical images. The dark current background is negligible in the case of {\it AstroSat}-UVIT \citep{Tandon2017}. Therefore the background for UVIT is dominated by astronomical backgrounds which are typically low in the order of $10^{-3}$ to $10^{-4}$ photons s$^{-1}$ arcsec$^{-2}$ \citep{Murthy2010}. As a result, the distribution of counts from the UVIT background is expected to follow the Poisson distribution due to its low counts. 

%{\bf A similar approach was adopted in the preparation of the GALEX source catalog \citep{Morrissey2007}}. 

Currently, there exist source catalogs for a few small regions of the sky \citep[e.g.,][]{Ashish2023, Leahy2020, Chayan2023} from UVIT observations. 
\citet{Leahy2020} considered 18 overlapping fields from the UVIT observation of M31 and produced a point source catalog of $\sim 75000$ objects detected in FUV or NUV filters. The survey's limiting magnitude (bin with the largest number of sources detected) was $\sim 23$ magnitude in the FUV CaF2 filter. \citet{Ashish2023}, considered three overlapping UVIT observations in the outskirts of SMC to produce the points source catalog of $\sim 11000$ sources detected across three FUV and four NUV filters. %Convolution with the Gaussian filter of full width at half maxima (FWHM) $\sim$ 1.5$^{\prime \prime}$ was carried out in both the study to enhance the source detection. Both the study utilized local peak-finding algorithms to detect the sources. 
However, these studies do not explicitly treat the Poisson distribution of the UVIT background to calculate background and source detection threshold.

In this work, we discuss an improved methodology for background calculation and source detection for AstroSat-UVIT observations, considering the Poisson distribution of the UVIT background and its importance in detecting the sources in the low flux limit. Details of the data used in this work are discussed in the next section. In section 3, the analysis and results will be discussed. Section 4 provides the discussion and conclusions.

\section{Data}

\begin{figure*}
\centering
\includegraphics[angle=0,height=6cm,width=16cm]{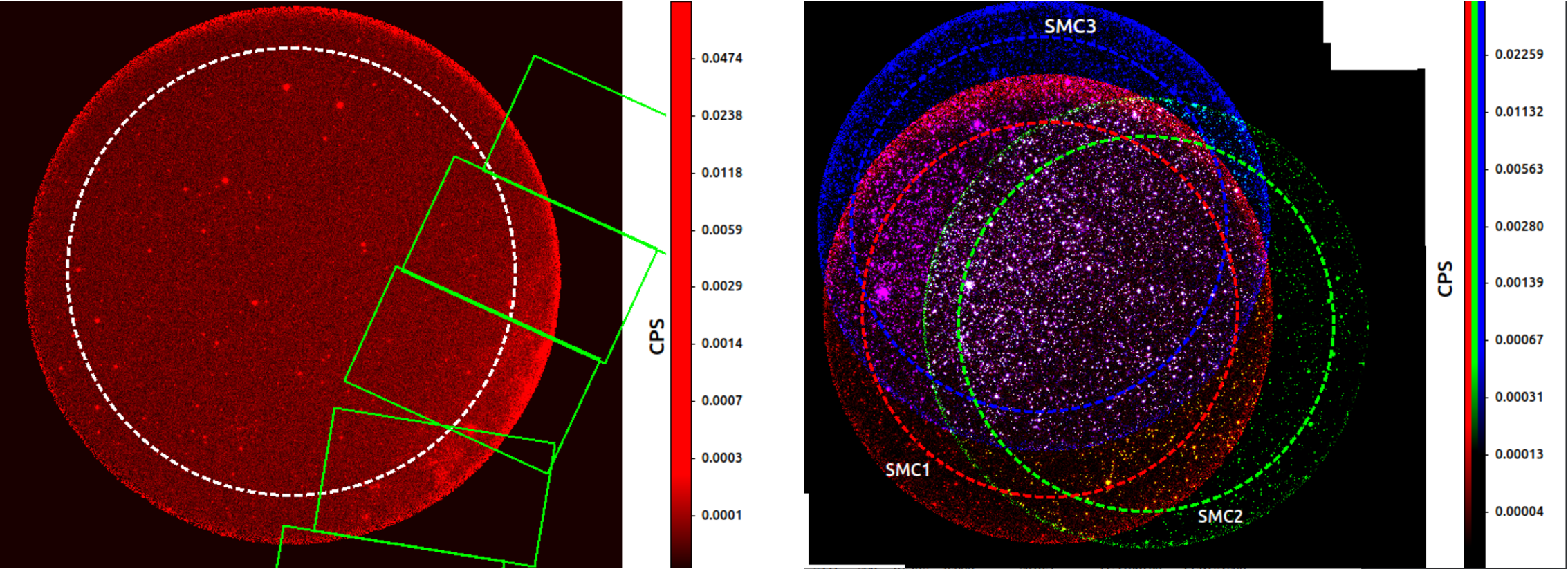}
 \caption{Regions used for comparison in this work. Left panel: UVIT observation of M31 Field 13 in NUVN2 filter. White dashed circles represent the central $12^{\prime}$ region, and green boxes represent the PHAT survey overlapping regions. Right panel: UVIT observations of three partially overlapping MC regions in the NUVB13 filter. Red, green, and blue correspond to SMC1, SMC2, and SMC3 fields, respectively. The inner dashed circle corresponds to the central $12^{\prime}$ regions of each field used in the current analysis. \label{Region_comp}}
   
\end{figure*}

UVIT onboard AstroSat telescope consists of two Ritchey-Chretien telescopes,  which observe in FUV (130-180 nm) and NUV (200-300 nm). It has a FOV of $14^{\prime}$ radius and a resolution of $\sim 1.2$-$1.5^{\prime\prime}$ with a plate-scale of $0.416$ arcsec \citep[][]{Tandon2017, Tandon2020}. We utilized publicly available Level-1 data from the Indian Space Science Data Center (ISSDC)\footnote{$https://astrobrowse.issdc.gov.in/astro\_archive/\\archive/Home.jsp$} and processed it through the CCDLAB \citep{Postma2017, Postma2021} pipeline to obtain the Level-2 images and exposure maps. The pipeline corrects for various spacecraft drift, field distortions, and flat field correction to provide the counts' image and corresponding exposure map \citep[see][for details]{Postma2017, Postma2021}. We also applied the cosmic ray correction by removing the frames 4$\sigma$ above the median frame counts using CCDLAB \citep[see][for details]{Postma2021,Chayan2023}. In this work, we have used M31 Field 13 and Small Magellanic Cloud (SMC) fields to demonstrate the methodology of the catalog construction. 

{\bf M31 Field 13:} 
%M31 Field has been analyzed by \citet{Leahy2020} for producing the source catalog using UVIT observation. 
UVIT observations cover a large part of the M31 galaxy with a series of observations (PI: D. Leahy). Among those fields, we chose M31 Field 13 for the current analysis. M31 Field 13 is comparatively less crowded and has considerable overlap with the deep Hubble Panchromatic Andromeda Treasury (PHAT) survey \citep{Williams2014}. PHAT survey is a catalog of $117$ million sources in the M31 field using the Hubble telescope. One of the filter observations from Hubble, F275W ($\lambda$ = $275$ nm) has a central wavelength very close to the NUVN2 filter ($\lambda$ = $279$ nm). In the F275W filter, the catalog is complete to $50\%$ in the Vega magnitude range of $\sim 24.41$ - $25.06$, depending on stellar density \citep{Williams2014}.
%This catalog's 50$\%$ completeness Vega magnitude is, on average, $\sim$ $24.41$ - $25.06$ (depending on stellar density) in the $F275W$ filter \citep{Williams2014}. 
This region provides a platform to compare our methodology with the existing study from \citet{Leahy2020} and deep PHAT survey. Most of the results here are illustrated with M31 Field-13 as an example field. The extended details of the observation of M31 Field 13 are available at \citet{Leahy2020}. In M31 Field 13, we utilized broad filter FUVCaF2 in FUV and NUVN2 filter in NUV (as it has a central wavelength close to PHAT F275W filter) and compared the results with \citet{Leahy2020, Leahy2021} (Hereafter called L20 and L21 respectively) and PHAT survey. 

{\bf SMC:} In SMC, we have utilized three overlapping fields with an offset of $6^{\prime}$ with each other. The same observations were used by \citet{Ashish2023} to produce the point source catalog in this region. More details of observation of the SMC field can be obtained at \citet{Ashish2023} (hereafter called as D23). In SMC, we chose two broad FUVBaF2 and NUVB13 filters for comparison with D23. As there is no deep Hubble survey in SMC, we compared our results with the optical Gaia survey \citep{Gaia2016,Gaia2018}.

The brief details of the UVIT observations utilized in the current work are provided in Table~\ref{obs_detail}. The UVIT image of M31 Field 13 with PHAT overlapping region and three overlapping regions of SMC is provided in Figure.~\ref{Region_comp}

\begin{table*}
	%\centering
	\caption{Details of UVIT observations utilized. \label{obs_detail}}
	
	\begin{tabular}{ccccccccc} % four columns, alignment for each
		\hline
		 Field & Filter & $\lambda_{\text{mean}}$\tablenotemark{\scriptsize{1}} & $\Delta\lambda$\tablenotemark{\scriptsize{1}} & ZP\tablenotemark{\scriptsize{1}} & Error &Exp. time (sec.) & Center RA & Center DEC\\
            &   &($\text{\AA}$) & ($\text{\AA}$) &  & in ZP\tablenotemark{\scriptsize{1}} & &  (in deg) & (in deg) \\
		\hline
            \hline
            M31 & FUVCaF2 (F148W) & $1481$& $500$& $18.097$& $0.010$& $4561$ & $11.58483$ & $41.56887$ \\
             & NUVN2 (N279N) & $2792$& $90$& $16.416$& $0.010$& $4345$ &$11.60549$ & $41.55920$\\
            \hline
            SMC1 &FUVBaF2 (F154W) &$1541$& $380$& $17.771$& $0.010$& $1895$& $17.36675$& $-71.30638$\\
             & NUVB13 (N245M) & $2447$& $280$& $18.452$& $0.005$&$1895$& $17.42300$& $-71.31109$ \\
            
             \hline
            SMC2 &FUVBaF2 (F154W) & $1541$& $380$& $17.771$& $0.010$&$1888$& $17.04458$& $-71.29326$ \\
             & NUVB13 (N245M)&$2447$& $280$& $18.452$& $0.005$&$1898$& $17.10057$& $-71.29837$ \\
             
             \hline
             SMC3 &FUVBaF2 (F154W) &$1541$& $380$& $17.771$& $0.010$&$1904$ & $17.38476$& $-71.39599$ \\
             & NUVB13 (N245M)&$2447$& $280$& $18.452$& $0.005$&$1902$ & $17.44120$ & $-71.40139$ \\
             
             \hline
             %\multicolumn{8}{l}{$^{1}$\citep{Tandon2020}}\\
	\end{tabular}
 \tablenotetext{1}{\citet{Tandon2020}}
\end{table*}

\section[12pt]{Analysis and Results} 
\subsection {Nature of the UVIT background}

\begin{figure}
\centering
\includegraphics[angle=0,height=6cm,width=8cm]{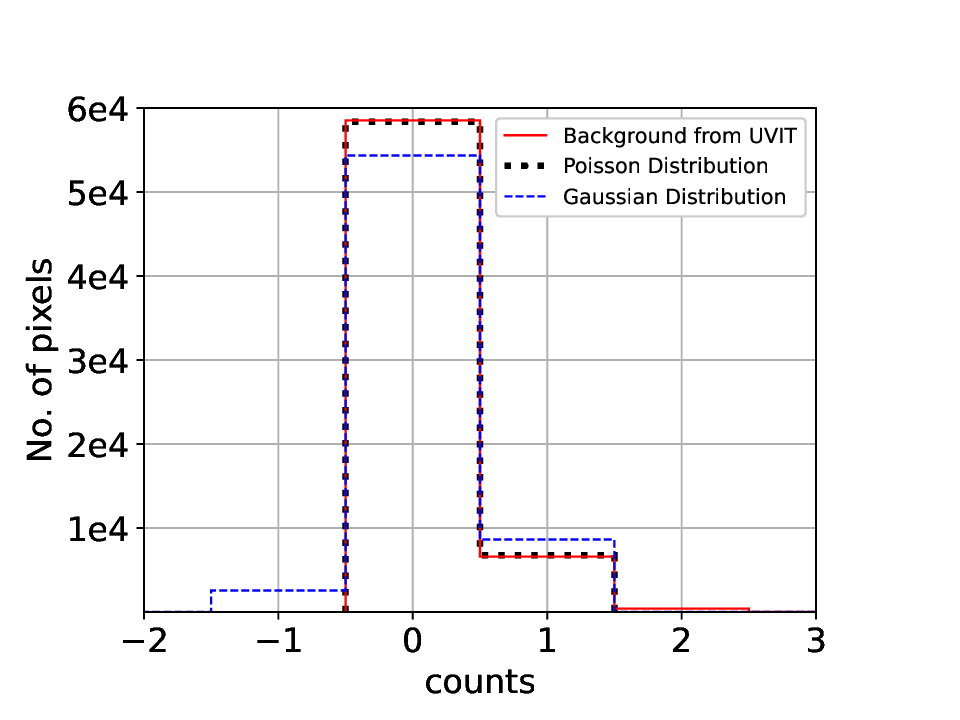}
\caption{Histogram of UVIT background distribution in $128\times128$ pixel source-free region compared with the expected Poisson and Gaussian distribution in the field of M31 Field 13 NUVN2 filter. \label{bg_hist_comp_poiss_vs_gauss}}

\end{figure}

To understand the nature of the UVIT background, we first chose a few source-free regions of $128\times128$ pixels 
in the UVIT fields. 
%and compared them with Gaussian and the Poisson distributions. 
In the M31 Field 13 NUVN2 filter, we chose twenty such regions and calculated the mean and standard deviations of the background counts in those regions. We compared the histogram of background counts and the expected number of counts from the Gaussian and Poisson distributions. It is observed that the background distribution closely follows the Poisson distribution. The histogram from one such region is provided in Figure~\ref{bg_hist_comp_poiss_vs_gauss}.  

%We carried out the $\chi^2$ test \citep{Bevington2003} to quantify the deviation from the Poisson distribution. 
We observed that the assumption of the Gaussian distribution of the observed background yields a very high reduced $\chi^2$ (from $\sim 48$-$278$), suggesting that the background deviates significantly from the Gaussian distribution. The background is well explained by the Poisson distribution with reduced-$\chi^2$ (degrees of freedom ({\it dof})=3 for most of the regions) ranging from $0.12$ to $2.1$ and the majority of them lie between the value of $0.5-1.5$ which corresponds to a p-value of $0.8$ to $0.2$ for {\it dof} of $3$. %value of $1.07$. 
%These reduced-$\chi^2$ values suggest no significant deviation from the Poisson distribution. 
Similar trends are observed in the SMC field and other filter observations of M31 (FUVCaF2).

\begin{figure}
\centering
\includegraphics[angle=0,height=6cm,width=8cm]{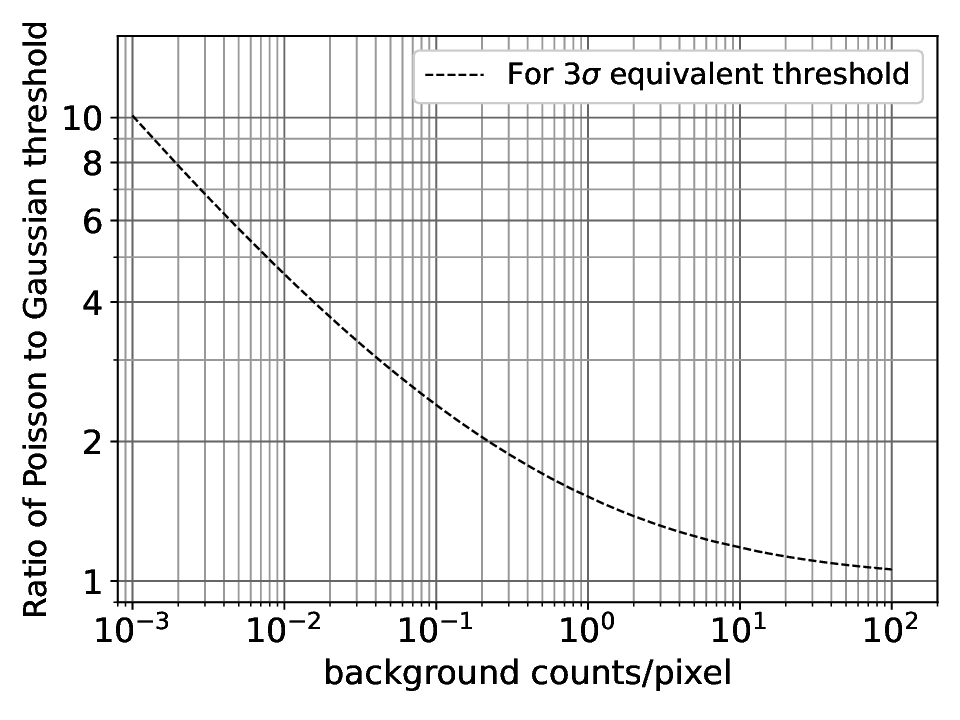}
\caption{Ratio of $3\sigma$ equivalent threshold from Gaussian and Poisson distribution as a function of background counts. \label{thresh_poiss_vs_gauss}}

\end{figure}

%The Poisson distribution probabilities converge to the Gaussian distribution if the background counts are high. However, 
For a Poisson distribution with mean ‘$\lambda$,’ the probability of getting a value above some value, k, is given by an incomplete gamma function, given below \citep[e.g.,][]{Nimai2013, Marsaglia1986}, where $\Gamma(x)$ is the Gamma function.

\begin{equation}
\centering
\label{poiss_prob}
P(X>x;\lambda) = \frac{\int_{0}^{\lambda}e^{-t}t^{x-1}dt}{\Gamma(x)}
%\dfrac{dF}{dE}=N %\left(\dfrac{E}{E_{b}}\right)^{-\alpha-\beta %\log{\left(\dfrac{E}{E_{b}}\right)}} 
\end{equation}
There is a significant deviation from the threshold calculated using the Gaussian and the Poisson distribution at low counts. The ratio of the $3\sigma$ equivalent threshold considering the Poisson distribution  (Probability of $\sim 99.73$$\%$) and the Gaussian distribution ($3\times$$\sqrt{\mbox{background}}$) as a function of background counts are provided in Figure~\ref{thresh_poiss_vs_gauss}.

\subsection{Astrometry Correction}
The astrometric solution, including any residual distortion for Level-2 images, is accomplished using the SCAMP \citep{Bertin2006} software. The Gaia-DR2 catalog \citep[][]{Gaia2016, Gaia2018, Arenou2018} is used as the reference catalog in the SCAMP. 
The SCAMP software interacts with SExtractor catalogs to provide astrometric solutions. The astrometric solutions are stored in the header of the fits file and are applicable to all the sources in the image. SExtrcator utilizes this header information to provide the WCS for the detected sources. The SCAMP makes no modifications to the image or pixel coordinate.
%SCAMP provides further correction for any residual distortion left in the astrometry. It provides the astrometry correction as the header without altering the image counts. It is also compatible with the SExtractor software used to detect the sources in this work. 
The root-mean-square (RMS) value of the difference in Gaia and UVIT position obtained from SCAMP is well within a pixel (0.2 to 0.6 pixel) of UVIT in all fields. 

%{ \st { The astrometric accuracy values of the matched Gaia source lie within a pixel of UVIT. The root-mean-square (RMS) value of the difference in Gaia and UVIT position obtained from SCAMP post astrometry is provided in Table. The observed values are well within a pixel scale of UVIT in all the fields.}}

%(0.4168$^{\prime\prime}$)

%This value also agrees with the expected scatter ($0.24$ arcsec) of L20 in the same field. 
%\begin{comment}
%\begin{table}
%	\centering
%	\caption{Comparison of astrometric accuracy in our comparison fields with Gaia. \label{t:Astrometry_std}}
	
%	\begin{tabular}{ccc} % four columns, alignment for each
%		\hline
%		Field & Filter & RMS of difference  \\
%                & &in  Gaia and UVIT \\
%                & & position (in arcsec)\\
%		\hline
%            \hline
%                M31 Field 13& FUVCaF2 &$0.2473$  \\
%                & NUVN2&$0.1354$ \\
%            \hline
%              SMC1 & FUVBaF2&$0.1246$ \\
%               & NUVB13 &$0.1773$ \\
%                \hline
%               SMC2 & FUVBaF2&$0.1619$ \\
%               & NUVB13 &$0.102$ \\
%                \hline
%               SMC3 & FUVBaF2&$0.197$ \\
%               & NUVB13 &$0.095$ \\
%               %& NUVN2&0.192$^2$ \\
%            \hline
%	\end{tabular}
%\end{table}
%\end{comment}

\subsection{Background and Threshold Calculation}
\subsubsection{Background Calculation:}
 
The background is calculated considering the Poisson distribution following a similar methodology, as applied in the GALEX \citep{Morrissey2007}. The steps followed in the calculation of the background are as follows: 
\begin{enumerate}
\item Initially, the pixels with exposure less than $20\%$ of the total exposure time are masked and not considered for further analysis. 
\item The count map is divided into blocks where the background is calculated. 
\item 
%{\st {The background is assumed to be constant within this block. The choice of block size is made so that it is large enough to accommodate most of the bright sources in the field, but small-scale variations in the background can still be observed. We considered the block size of $128\times128$ pixels for background calculation.}}
The background is assumed to be constant within this block. We used the block size of $128\times128$ pixels for background calculation. This choice of block size is made so that it is large enough to accommodate most of the bright sources in the field but provides variations in the background on scales greater than the block size.
\item 
%{\st {The background in each block is calculated iteratively after removing the pixels above equivalent three sigma (significance level of $1.35\times 10^{-3}$) as calculated by equation~1.}} 
To calculate the background in each block, pixels with count rates above equivalent to three sigma (using equation~\ref{poiss_prob}) are removed iteratively until the mean count rate in the block ceases to vary. Further, since the background calculated in a block of $128\times128$ pixels could be overestimated due to the presence of an extended/large number of sources, we applied a $3\times3$ median filter. The background for a given block is taken as the median of 9 blocks ($3\times3$ array centered around the block). 

\item The background obtained for each block is assigned to the central pixel of the corresponding block. By sequential horizontal and vertical interpolation of these central pixel count rates (bi-linear interpolation using {\it scipy.interpolate.interp2d} routine in {\it Python}), the background count rates for the remaining pixels were calculated.
%\st {To avoid overestimating the background by bright extended or crowded sources, a median filter of $3\times3$ block is applied.} 
\item 
%\st{The background in each mesh is interpolated using bi-linear interpolation to obtain the background of the same resolution as the input image.} 
We subtract the background from the level-2 image and divide it by the detection threshold map. This image is called a ``detection image.'' The pixels above the detection threshold have values greater than unity in this image. The source detection is carried out on the detection image.

\item SExtractor software \citep{Bertin1996} was used for source detection. It is run twice (with parameter values as provided in Table 2) to produce the source list for each observation field. During the first run, SExtractor’s segmentation map is used to generate a mask file for the second iteration. The segmentation map has non-zero values for pixels above the detection threshold around the sources, and these pixels are masked. The background and threshold are calculated again by excluding masked regions. The SExtractor is rerun to yield the final source list. 
%\st {To take care of bright and extended sources in the field, background, and threshold calculation and source detection is carried out twice. The sources detected in the first iteration are masked using a segmentation map from the SExtractor  (Bertin \& Arnouts 1996) (refer Section. 3.4). The background, threshold calculation, and source detection are carried out again to yield the final catalog.}
\end{enumerate}

%\st {We further considered the 200 randomly selected positions in each background-subtracted image. We calculated the aperture flux in a 3-pixel radius for these positions. The zero count rate was observed to fall within one standard deviation of the mean.} 
We further considered the 200 randomly selected source-free positions in each background-subtracted image. We calculated the aperture flux in a 3-pixel radius for these positions. We observe that the mean of the residuals falls near zero, within three standard errors. This result suggests that the assumption of the Poisson nature of the background is essential for UVIT source detection. 

%{\st{As an example, Figure shows a histogram of residuals in the randomly selected regions of the M31 Field 13 (NUVN2 filter).}}

%{As an example, Figure \ref{bg_sf_region} shows a histogram of residuals in the randomly selected regions of the M31 Field 13 (NUVN2 filter).

%\begin{comment}
%\begin{figure}
%\centering
%\includegraphics[angle=0,height=6cm,width=8cm]{hist_source_free_region_cps.eps}
%\caption{Histograms of count rate before and after background subtraction (residuals) in M31 Field 13 (NUVN2 filter).  As desired, the zero count rate falls within one standard deviation of the mean of residuals. \label{bg_sf_region}}

%\end{figure}
%\end{commnet}

\subsubsection{Threshold Calculation:}
In SExtractor (see Section. \ref{src_det}), the  
%\st {background-subtracted, threshold-divided} 
detection image is convolved with the point spread function (psf) to maximize the signal-to-noise ratio (SNR) \citep[e.g.,][]{Bijaoui1970, Irwin1985,  Akhlaghi2015}. However, as convolution suppresses the noise variance, the source detection threshold has to be modified accordingly. For white noise and the Gaussian convolution kernel, the variance reduces by a factor of $4 \pi \sigma^{2}$, where $\sigma$ is the standard deviation of the Gaussian convolution kernel \citep[e.g.,][]{Buades2010}. For the Poisson distribution, one of the methods for calculation of the probability post-convolution can be given by Fay and Feuer approximation \citep[e.g.,][]{Stewart2006}, according to which the probability of obtaining value greater than $x$ post-convolution ($P^{\prime}(X\geq x;\lambda)$)is given by \citep{Stewart2006},

\begin{equation}
\centering
\label{fay_feuer_approx}
P^{\prime}(X\geq x;\lambda) = P(X\geq \frac{x}{w};\frac{\lambda}{w})  %\frac{\int_{0}^{\lambda^{\prime}}e^{-\lambda^{\prime}}\lambda^{\prime^{x-1}}dx}{\Gamma(x)}
%\dfrac{dF}{dE}=N %\left(\dfrac{E}{E_{b}}\right)^{-\alpha-\beta %\log{\left(\dfrac{E}{E_{b}}\right)}} 
\end{equation}

Where $w$ is the weight factor due to convolution.

The weight factor ($w$), for the Gaussian convolution kernel, can be given by \citep{Stewart2006}, 
\begin{equation}
\centering
\label{weight_fay_feuer}
w=\frac{1}{4 \pi \sigma^{2}}
 %\frac{\int_{0}^{\lambda^{\prime}}e^{-\lambda^{\prime}}\lambda^{\prime^{x-1}}dx}{\Gamma(x)}
%\dfrac{dF}{dE}=N %\left(\dfrac{E}{E_{b}}\right)^{-\alpha-\beta %\log{\left(\dfrac{E}{E_{b}}\right)}} 
\end{equation}

For a high background regime, when the Poisson distribution converges to the Gaussian distribution, the resulting weight factor (reduction in variance) tends to the expected factor of $4 \pi \sigma^{2}$ of the Gaussian distribution. We utilized equation~(\ref{fay_feuer_approx}) to calculate the source detection threshold in this work. % , and the Gaussian convolution filter of FWHM $=$ 3 pixels was used in the SExtractor. We considered a 4-$\sigma$ equivalent threshold for the detection of the sources in our method.
%However, this prescription underestimates the threshold calculated if the background count is less than $0.1$ and if the probability is less than $10^{-2}$ \citep{Stewart2006}. Also, the prescription does not consider the correlation between the pixels post-convolution. Involving the non-Gaussian PSF and no standard methods for the calculation of the false detection probability, we also utilized the simulations and comparison with other catalogs to set up the detection threshold.

\subsection{Source detection and photometry}\label{src_det}

 %Since the background and threshold were calculated considering Poisson distribution, 
 The  
 %\st {background subtracted, threshold divided} 
 detection image is provided as the input for detecting the sources in SExtractor (Version {\it 2.25.0}). The Gaussian convolution filter with full width at half maximum (FWHM) equal to $\sim 3$ ($\sim$ psf of UVIT) pixels is used. The DETECT\_MINAREA is chosen as 9 (Nearly the number of pixels within the box of 1 FWHM). The values of the parameters used for background calculation and source detection are provided in Table~\ref{SE_param}. The photometry is carried out on the background-subtracted image utilizing the aperture photometry technique. The source fluxes are calculated in the aperture of $3$ pixels centered at each source position using SExtractor. The aperture and saturation corrections are applied utilizing the methodology provided in \citet{Tandon2020}. Due to distortion beyond the central $12^{\prime}$ region \citep{Tandon2020}, we limited our source list to the sources detected in the radius of central $12^{\prime}$ of the field of view (FOV). 
 
 In M31 Field 13, we considered the region within 12$^{\prime}$ radius of the center of UVIT NUVN2 field and PHAT survey \citep{Dalcanton2012, Williams2014} overlapping region, which allows us to independently compare sources detected in this work with deeper PHAT survey.

For SMC, we constructed the source list of sources in each of the three fields. We selected sources within $12^{\prime}$ of each field. Since the fields are partially overlapping, we used {\it Internal Match}  routine of {\it TOPCAT}\footnote{https://www.star.bris.ac.uk/$\sim$mbt/topcat/} software \citep{Taylor2005}  to remove the multiple entries in the overlapping region. We retained only the brightest source among those lying within $1.5^{\prime\prime}$ of each other. This radius is optimal at which the change in the number of sources, as a function of the matching radius, reaches a minimum.
%Beyond this radius, the change in the number of sources increases again with the match radius indicating the loss of real sources. We utilized the   for removing the possible multiple entries.} %We also selected the sources from D23 that lie within our analysis region.

We detected $194$ sources in FUVCaF2 and $141$ sources in the NUVN2 filter in this comparison region of M31 and $9445$ sources in FUVBaF2 and $15354$ sources in the NUVB13 field in the comparison region of SMC.

\begin{table}
	\centering
	\caption{Parameters for source detection and analysis. \label{SE_param}}
	
	\begin{tabular}{ll} % four columns, alignment for each
		\hline
		Parameter & Value \\
		\hline
            \multicolumn{2}{c}{\bf Background and threshold calculation parameter}\\
            \hline
            background mesh & $128\times 128$ pixels\\
            Threshold & $4\sigma$ equivalent\\
            Weight factor (To account  & $20.4$\\
            for convolution) &\\
            Median filter & $3\times3$ meshes\\
            Mask & Segmentation map \\
            & from SExtractor\\
            \hline
            \multicolumn{2}{c}{\bf SExtractor parameter}\\
            \hline
            DETECT$\_$MINAREA & $9$ \\
            THRESH$\_$TYPE & ABSOLUTE \\
            DETECT$\_$THRESH & $1$ \\
            Convolution & Gaussian filter \\
             &of FWHM 3 pixels\\
             DEBLEND$\_$MINCONT & $0.005$\\
             DEBLEND$\_$NTHRESH & $64$ \\
            PHOT$\_$APERTURES & $6$ \\
            BACK$\_$TYPE & MANUAL \\
            BACK$\_$VALUE & $0.0$ \\
            \hline
	\end{tabular}
\end{table}

\subsection{Catalog Parameters}
%The catalog for different filters is provided separately. 
The catalog contains the source ID containing the detail of the filter in which it is detected and the position of the source in the format, ``UVIT+$<$filter name$>$+J+$<$RA in degrees$>$+$<$DEC in degrees$>$''. The RA and DEC of the sources are calculated by the SExtractor. Due to the small scatter, we adopted the windowed position from the SExtractor. The catalog also provides the magnitude and magnitude error of the sources in different filters. The first few lines of the table are provided here (Table~\ref{t:Catalog}) for reference. The entire catalog is available as supplementary material in the online version of the journal. 
\begin{table*}
	\centering
	\caption{First few lines of the catalog of UV sources in SMC and M31 Field 13 PHAT survey overlapping region. Table~\ref{t:Catalog} is published in its entirety in the machine-readable format. A portion is shown here for guidance regarding its form and content. \label{t:Catalog}}

	\begin{tabular}{cccccc} % four columns, alignment for each
		\hline
		Source &Field & RA & DEC  & AB Mag   & Mag err  \\
  ID&& (degrees)& (degrees)& & \\
  \hline
UVITF148WJ11.53065+41.74304& M31F13& $11.53065$&  $41.74304$& $21.83$& $0.124$\\ 
UVITF148WJ11.5238+41.72935&  M31F13& $11.52380$&  $41.72935$& $22.62$& $0.182$ \\
UVITF148WJ11.51992+41.74682& M31F13& $11.51992$&  $41.74682$& $22.59$& $0.182$\\ 
UVITF148WJ11.51934+41.74331& M31F13& $11.51934$&  $41.74331$& $20.15$& $0.056$ \\
UVITF148WJ11.51551+41.72488& M31F13& $11.51551$&  $41.72488$& $23.16$& $0.242$\\ 
UVITF148WJ11.51506+41.73659& M31F13& $11.51506$&  $41.73659$& $23.05$& $0.231$ \\
UVITF148WJ11.51444+41.63572& M31F13& $11.51444$&  $41.63572$& $22.81$& $0.200$\\ 
UVITF148WJ11.51024+41.7191&  M31F13& $11.51024$&  $41.71910$& $22.61$& $0.181$ \\
UVITF148WJ11.50738+41.73452& M31F13& $11.50738$&  $41.73452$& $23.46$& $0.293$ \\
UVITF148WJ11.50601+41.74007& M31F13& $11.50601$&  $41.74007$& $22.54$& $0.177$ \\
.& . &. &. &. &. \\ 
.& . &. &. &. &. \\
.&  . &. &. &. &. \\
 
            \hline
	\end{tabular}
\end{table*}

\section{Discussions and Conclusions}
\subsection{Comparison with other catalogs}\label{Comp_cat}
\begin{figure*}
\centering
\includegraphics[angle=0,height=8cm,width=16cm]{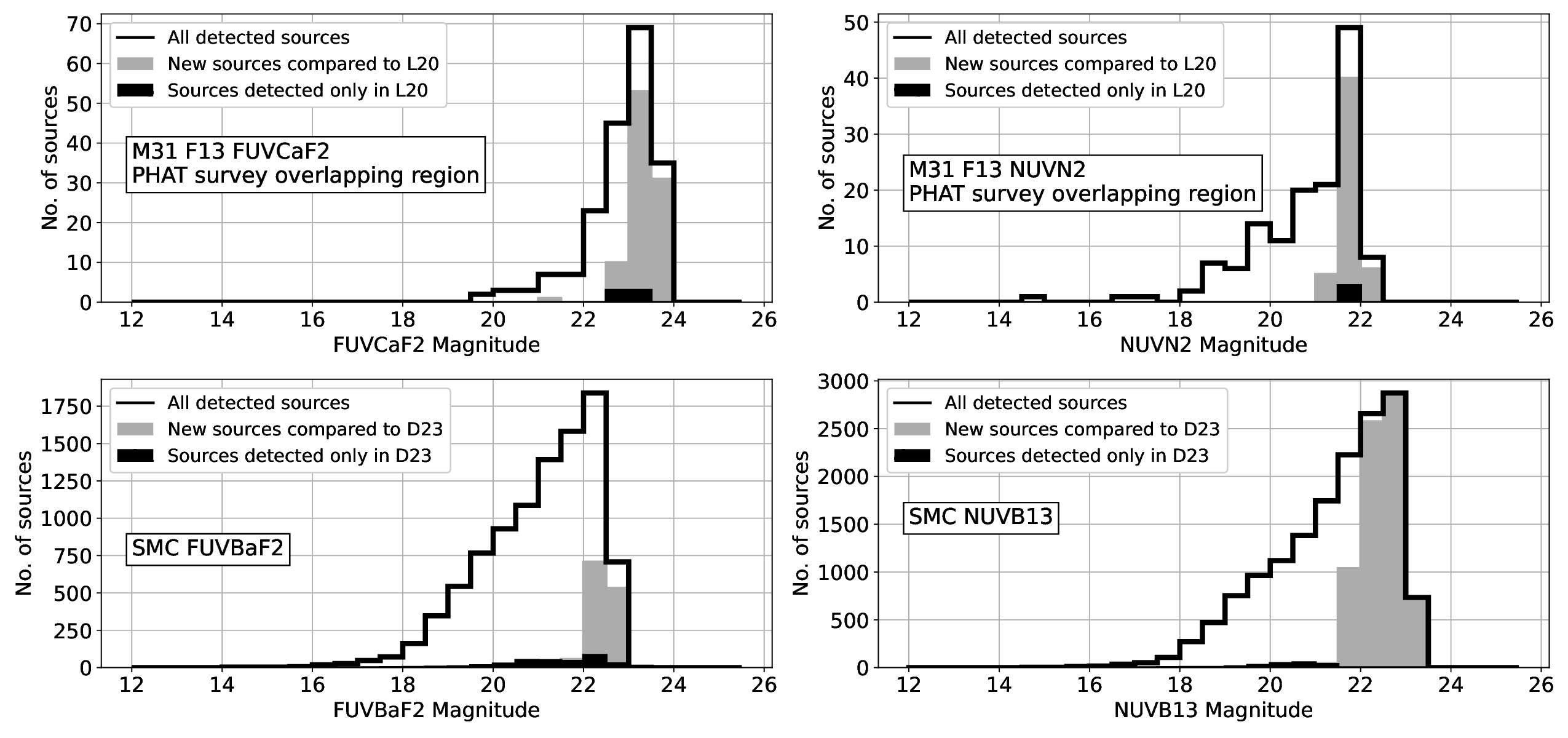}
\caption{Magnitude distribution of all detected and new sources from this work compared to L20 and D23. Also shown are sources detected only in previous studies in M31 Field 13 and SMC. Top left panel: M31 Field 13 PHAT survey overlapping region in FUVCaF2 filter compared with L20 \citep{Leahy2020} Top right panel: M31 Field 13 PHAT survey overlapping region in NUVN2 filter compared with L20. Bottom left panel: SMC in FUVBaF2 filter compared with D23 \citep{Ashish2023} Bottom right panel: SMC in NUVB13 filter compared with D23. \label{Mag_distribution}} 

\end{figure*}

\begin{table*}

  %\begin{threeparttable}
%\aboverulesep=0ex % Solution part 1 of 3
 %  \belowrulesep=0ex % Solution part 1 of 3
	%\centering
	\caption{Details of the comparison of detected sources in this work with other studies in M31 Field 13 and SMC using UVIT observations and other waveband catalogs. \label{t:Source_comp}}

\begin{tabular}{|l|l|c|c|c|c|}

%\begin{tabular}[t]{|p{5em}| p{10em}| >{\centering}p{6em}| >{\centering}p{6em}| >{\centering}p{6em}| >{\centering\arraybackslash}p{6em}|} % four columns, alignment for each
 %\multicolumn{5}{c}{M31 Field 13 PHAT survey overlapping region}\\
 %\toprule
 \hline
 \rule{0pt}{1.1EM}% Solution part 2 of 3 (% is required)
  Field & & \multicolumn{2}{c|}{FUV\tablenotemark{\scriptsize {a}}} & \multicolumn{2}{c|}{NUV\tablenotemark{\scriptsize {b}}} \\
  \hline
  \hline
 
   \rule{0pt}{1.1EM}
M31  &Total Number of  &\multicolumn{2}{c|}{$194$ ($103$)} &\multicolumn{2}{c|}{$141$ ($93$)}\\
   Field 13\tablenotemark{\scriptsize {c}}& sources detected\tablenotemark{\scriptsize {d}} &\multicolumn{2}{c|}{} &\multicolumn{2}{c|}{}\\ 
   \cline{2-6}
 %   \cmidrule(){2-6}
     \rule{0pt}{1.1EM}
   &Counterpart from &\multicolumn{2}{c|}{$97$} &\multicolumn{2}{c|}{$90$}\\
   & previous work\tablenotemark{\scriptsize {e}} &\multicolumn{2}{c|}{} &\multicolumn{2}{c|}{}\\
   \cline{2-6}
   %\cmidrule(){2-6}
    \rule{0pt}{1.1EM}
&New sources\tablenotemark{\scriptsize {f}} &\multicolumn{2}{c|}{$95$} &\multicolumn{2}{c|}{$51$}\\
&&\multicolumn{2}{c|}{ } &\multicolumn{2}{c|}{ }\\
\cline{3-6}
%\cmidrule(){3-6}
% \rule{0pt}{1.1EM}
& &Sources with &No counterpart &Sources with&No counterpart \\
& &counterpart in& in PHAT& counterpart&in PHAT\\
& &in PHAT\tablenotemark{\scriptsize {g}}&& in PHAT\tablenotemark{\scriptsize {g}}&\\
\cline{3-6}
%\cmidrule(){3-6}
 \rule{0pt}{1.1EM}
& & $81$& $14$&$42$ &$9$\\
\hline
\hline
%  \midrule
%  \midrule
 \rule{0pt}{1.1EM}
  SMC &Total Number of  &\multicolumn{2}{c|}{$9539$ ($8563$)} &\multicolumn{2}{c|}{$15433$ ($8360$)}\\
 
   & sources detected\tablenotemark{\scriptsize {d}} &\multicolumn{2}{c|}{} &\multicolumn{2}{c|}{}\\ 
   \cline{2-6}
%    \cmidrule(){2-6}
     \rule{0pt}{1.1EM}
   &Counterpart from  &\multicolumn{2}{c|}{$8257$} &\multicolumn{2}{c|}{$8181$}\\
   & previous work\tablenotemark{\scriptsize {e}} &\multicolumn{2}{c|}{} &\multicolumn{2}{c|}{}\\
   \cline{2-6}
   %\cmidrule(){2-6}
    \rule{0pt}{1.1EM}
&New sources\tablenotemark{\scriptsize {f}} &\multicolumn{2}{c|}{$1304$} &\multicolumn{2}{c|}{$7270$}\\
&&\multicolumn{2}{c|}{ } &\multicolumn{2}{c|}{ }\\
\cline{3-6}
%\cmidrule(){3-6}
 \rule{0pt}{1.1EM}
& &Sources with &No counterpart  &Sources with  &No counterpart \\
& &counterpart &in Gaia &counterpart &in Gaia\\
& &in Gaia& &in Gaia &\\
\cline{3-6}
%\cmidrule(){3-6}
 \rule{0pt}{1.1EM}
& & $1152$& $152$&$4664$ &$2606$\\
\hline
% \bottomrule
\multicolumn{6}{l}{ } \\
	\end{tabular}

\tablenotetext{\scriptsize{a}}{FUVCaF2 in M31 and FUVBaF2 in SMC.}

\tablenotetext{\scriptsize {b}}{NUVN2 in M31 and NUVB13 in SMC.}

\tablenotetext{\scriptsize {c}}{In the Panchromatic Hubble Andromeda Treasury (PHAT) \citep{Williams2014} survey overlapping region.}

\tablenotetext{\scriptsize {d}}{In bracket provided the number of the sources detected in previous work; L20 \citep{Leahy2020} for M31 and D23 \citep{Ashish2023} for SMC in the comparison region of the given filter.}

\tablenotetext{\scriptsize {e}}{Number of sources matched within $1.77^{\prime\prime}$ from previous work (L20 for M31 and D23 for SMC) with this work.}

\tablenotetext{\scriptsize {f}}{Due to source confusion, the sources having multiple counterparts with the previous work are excluded.}

\tablenotetext{\scriptsize {g}}{A magnitude cut-off of $21.41$ and $23.23$ was applied in the PHAT survey catalog for comparison of sources in NUV and FUV filters, respectively.}
 %\begin{tablenotes}
%      \small
 %     \item $^{a}$ FUVCaF2 in M31 and FUVBaF2 in SMC.
 %     \item $^{b}$ NUVN2 in M31 and NUVB13 in SMC.
%      \item $^{c}$ In the Panchromatic Hubble Andromeda Treasury (PHAT) \citep{Williams2014} survey overlapping region.
 %     \item $^{d}$ In bracket provided the number of the sources detected in previous work; L20 \citep{Leahy2020} for M31 and D23 \citep{Ashish2023} for SMC in the comparison region of the given filter.
%      \item $^{e}$ Number of sources matched within 1.77$^{\prime\prime}$ from previous work (L20 for M31 and D23 for SMC) with this work.
%      \item$^{f}$ Due to source confusion, the sources having multiple counterparts with the previous work are excluded.
%      \item $^{g}$ A magnitude cut-off of 21.41 and 23.23 was applied in the PHAT survey catalog for comparison of sources in NUV and FUV filters, respectively.
%    \end{tablenotes}
%  \end{threeparttable}
\end{table*}

Our approach towards detecting UV sources from UVIT observations treats the background as a Poisson distribution. We compared the sources detected in this work with previously published UVIT catalogs (L20 and D23). We used a matching radius of $1.77$$^{\prime \prime}$ ($\sim \sqrt{2}$ times the FWHM of UVIT psf).
However, when searching for counterparts using catalogs with very small source position uncertainty (PHAT and Gaia), we adopted a fixed matching radius of $1.25^{\prime\prime}$ ($\sim$ FWHM of UVIT). The summary results of the comparison with the L20 and D23 and other waveband catalogs are provided in Table~\ref{t:Source_comp}. The magnitude distribution of all sources detected in this work, along with new sources and sources only in the previous study compared to the L20 and D23, are provided in Figure~\ref{Mag_distribution}.

\subsection{Consistancy with earlier work}
In M31 Field 13,  $\sim 94\%$ of L20 sources in FUVCaF2 and $\sim 97\%$ in NUVN2 are present in the catalog of this work. Flux values for the spatially-matched sources are also consistent with the recent results of L20 and L21. Moreover, compared to L20, we have 51 ($\sim 36\%$ of our sources) new sources in NUVN2 and 95 sources ($\sim 49\%$ of our sources) in the FUVCaF2 filter. 

In SMC, for comparison with D23, we selected sources within a radius of $12^\prime$ of each field using the centers provided in D23. Nearly $96\%$ in FUVBaF2 ($\sim 98\%$ in NUVB13) of D23 sources are present in our source list. 
%If we decrease the deblending parameter DEBLEND$\_$MINCOUNT to 0.001, then an additional $\sim$ 1$\%$ of the sources of D23 had a counterpart in our catalog. However, the reduction of the deblending parameter deblended some fraction of sources that are separated even within one pixel of each other, which is unlikely to be a real source. Therefore, we decided on the default value of 0.005, where the separation between two nearby sources was at least 1 FWHM. 
Additionally, we have $1304$ ($\sim 14\%$ of our sources) sources in FUV and $7270$ ($\sim 47\%$ of our sources) new sources in NUV that did not have a counterpart in D23. The calculated magnitudes are consistent with those provided by \citet{ErratumDevaraj2023} in SMC. 

%The fact that most sources from previous studies are present in our catalog suggests that our catalog is consistent with the previous studies using UVIT observations.

%We used the magnitudes provided in \citet{ErratumDevaraj2023} for the magnitude comparison in SMC. The 
%D23 had a magnitude offset in their magnitude calculation (via private communication). Therefore, for comparing the magnitude with D23, we calculated this offset, and it was found to be $0.427\pm0.001$ in FUVBaF2 and $0.327\pm0.003$ in NUVB13 filter. This offset was added to the D23 magnitude for the comparison of the source magnitude.

%These comparisons suggest that we could satisfactorily reproduce the existing catalogs.  \&\ref{t:Source_comp_SMC}. 
\subsection{Confidence in the new sources detected}

In the case of new sources detected in this work, we compared them with existing catalogs in other wavebands to rule out any possible artifact/false detection. 

{\bf M31 Field 13:} We compared new sources from this work with the PHAT survey catalog\citep{Williams2014}.
%, which includes the NUV filter F275W, having a central wavelength close to that of the NUVN2 filter. 
For higher reliability, we only considered those PHAT sources with a `True' status for the F275W band. There are nearly $35000$ PHAT sources in our comparison region. Hence, the probability of an accidental match is high (up to $35$\% as calculated from random synthetic sources placed within the field of view).

To address this, we imposed a magnitude cut-off in the PHAT survey. The faintest source detected in the UVIT NUVN2 filter is $22.19$. Conversion of AB magnitude to Vega magnitude requires a subtraction of $1.48$ from NUVN2 magnitude (L21).  We considered an additional $2\sigma$ error ($0.7$ mag) of the faintest sources detected to account for the variation in the magnitude in the non-simultaneous UVIT and HST observation. With this criteria, we applied a magnitude cut-off of $21.41$ in the PHAT survey to compare with the UVIT NUVN2 filter. 

In the UVIT FUVCaF2 filter, the faintest source detected was of magnitude $23.83$. To compare the FUVCaF2 source list with the PHAT survey, we considered the color differences, i.e., NUVN2-FUVCaF2 magnitude between UVIT FUV and NUV sources. To obtain the color, we selected sources having only one counterpart between the UVIT FUV and NUV catalogs. We then calculated the interquartile range (IQR) of the color for these sources. We removed the sources having color $1.5\times$ IQR below the first quartile (Q1) and above the third quartile (Q3) to remove the possible outliers. The maximum color of $0.211$ mag was observed for the remaining sources. With this color difference, along with a $2\sigma$ error of the faintest source ($0.67$ mag) and conversion from AB to Vega magnitude ($-1.48$ mag), a magnitude cut-off of $23.23$ mag was applied in the PHAT survey for comparison with the FUVCaF2 filter. This magnitude cut-offs in FUV and NUV limits the accidental match probability to $1$-$2$\% in NUVN2 ($\sim$ $2$-$3$\% in FUVCaF2) calculated from synthetic sources placed at random positions in the field.

 Out of our $51$ new sources detected in the NUVN2 filter, $\sim 82\%$ are present in the PHAT survey within $1.25^{\prime\prime}$. In the FUVCaF2 filter, we have $95$ new sources, of which $\sim 85\%$ are present in the PHAT catalog. The existence of counterparts for most of our sources in the magnitude-limited PHAT survey provides confidence in these new sources. 

%\begin{comment}
\begin{figure}
\centering
\includegraphics[angle=0,height=6cm,width=8cm]{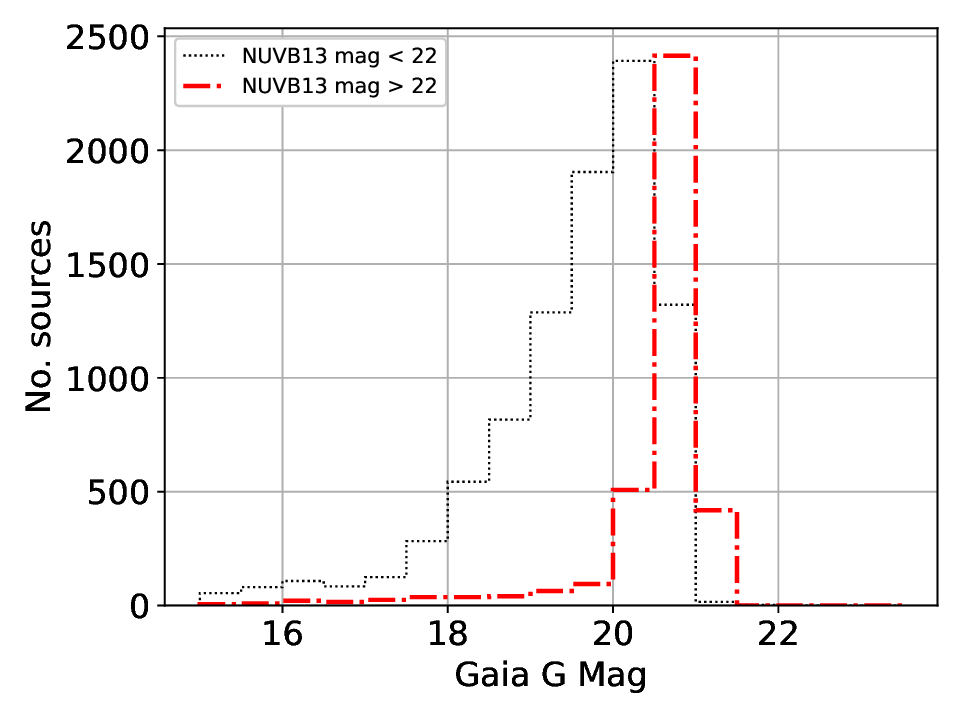}
\caption{Gaia G magnitude in NUVB13 filter of SMC field for 
 sources NUVB13 magnitude less than $22$ (dotted) and greater than $22$ (dash-dotted). \label{GAIA_NUV_B13_mag}}

\end{figure}
%\end{comment}
\begin{figure}[htbp]
\centering
\includegraphics[angle=0,height=8cm,width=8cm]{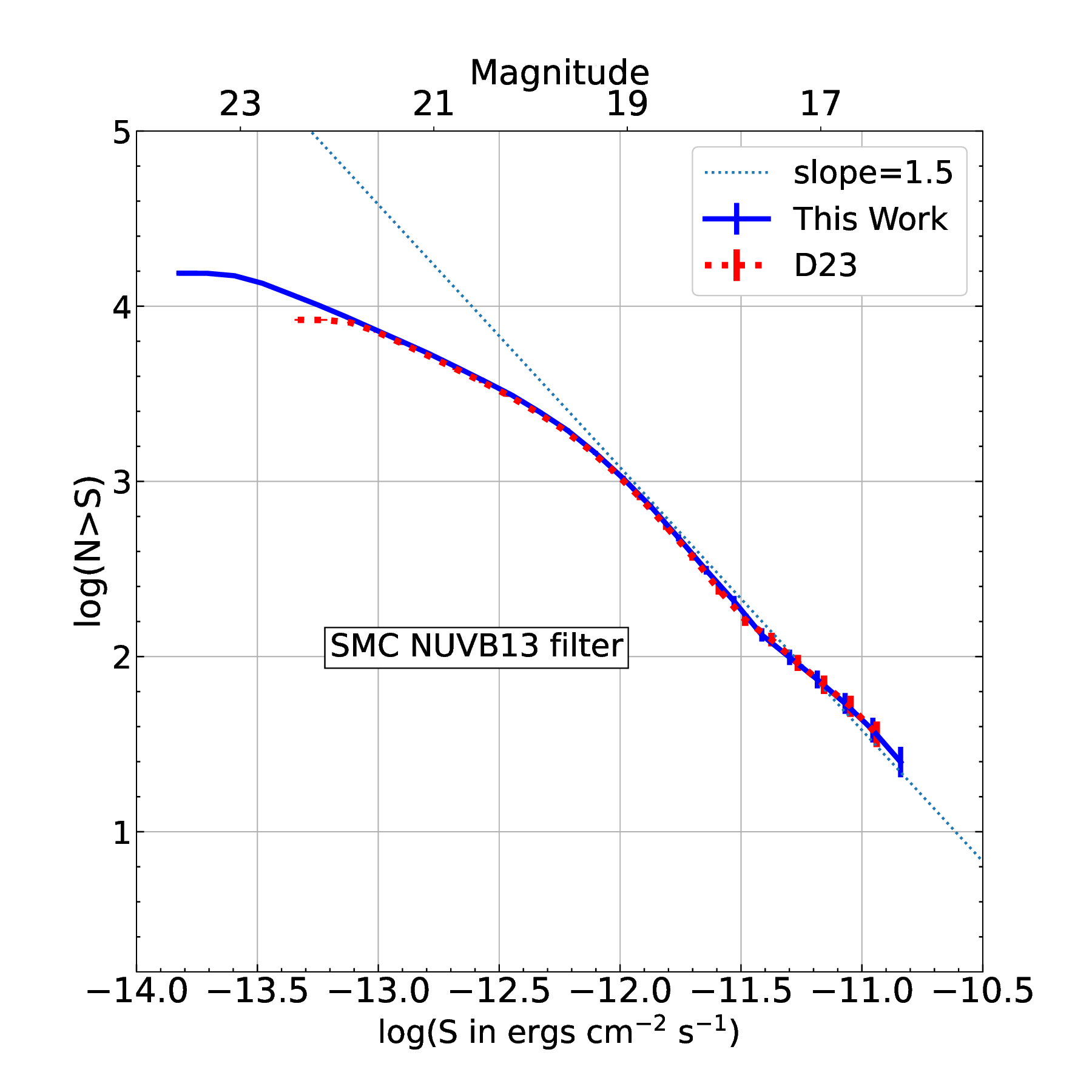}
\caption{Log(N) vs. Log(S) plot in the NUVB13 filter of the SMC field for sources detected in this work and sources detected in D23 \citep{Ashish2023}. A straight line of slope $1.5$ is plotted as a reference. The improvement in the identification of fainter sources is clearly visible. \label{lognlogs} }

\end{figure}

%\begin{comment}
%\begin{figure*}[htbp]
%\centering
%\includegraphics[angle=0,height=7cm,width=16cm]{Source_distribution_histogram.eps}
%\caption{Histogram of radial distribution of sourced having %counterpart in the existing UVIT studies compared with new sources %detected in this work. Left panel: In M31 Field 13 compared with L20 %\citep{Leahy2020}. Right panel: In SMC compared with D23 %\citep{Ashish2023}. \label{Position_distribution}}

%\end{figure*}
%\end{comment}

 {\bf SMC:} No extensive, deep Hubble survey is available for the SMC field. Hence, we compared the new sources with the optical Gaia-DR3 survey catalog \citep{Gaia2016,GAIA2023j} to look for counterparts. 
 
 In SMC, $\sim 14\%$, $47\%$ of our sources are new in FUVBaF2 and NUVB13, respectively, compared to D23. We noticed that a significant number of new sources detected in this work are present in the Gaia catalog ($\sim 88\%$ in BaF2, $\sim$ $64\%$ in NUVB13). 
 %We observed $\sim$ 90 \% in BaF2, $\sim$ $65\%$ in NUVB13 of our additional sources are present in the GAIA catalog. T

The non-existence of counterparts for a fraction of sources in the NUVB13 could be attributed to the high sensitivity of this filter (ZP = $18.452$ mag). Notably, $\sim 98.2\%$ of our new sources without Gaia counterparts have a magnitude of 22 mag or above in the NUVB13 filter. We further compared the Gaia magnitudes of high NUVB13 magnitude ($\geq 22$ mag) sources having Gaia counterparts and observed that they consistently have high Gaia magnitudes as seen in Figure~\ref{GAIA_NUV_B13_mag}. This observation supports the idea that the non-detection of a good fraction of our extra sources in Gaia is likely due to incompleteness of the Gaia survey beyond $18$ mag \footnote{https://gea.esac.esa.int/archive/documentation/GDR3/\\ Catalogue$\_$consolidation/chap$\_$cu9val/sec$\_$cu9val$\_$943/\\ssec$\_$cu9val$\_$943$\_$star$\_$density.html}. 

 The NUVB13 filter is nearly two magnitudes more sensitive than the NUVN2 filter. However, considering that the exposure time in NUVB13 for SMC observations is approximately $0.4$ times that of NUVN2 for M31 observations, we can expect the NUVB13 observations of SMC to be nearly 0.5-1 magnitude more sensitive. Indeed, we noticed that the magnitude distribution of NUVB13 sources in SMC peaks at approximately one magnitude fainter than those detected using the NUVN2 filter in M31 (as depicted in the top and bottom right panels of Figure~\ref{Mag_distribution}). This comparison suggests that the NUVB13 filter is capable of reaching a fainter magnitude limit, enabling the detection of fainter sources in the SMC.  

 The observed log(N) vs. log(S) plot in the NUVB13 filter of the SMC field is provided in Figure~\ref{lognlogs}. This plot clearly demonstrates that our method is efficient in detecting fainter sources. Different slopes are observed in low- and high-magnitude regimes. Two such slopes are also observed in the GALEX catalog sources \citep{Bianchi2017}. The slope is consistent with $1.5$ for brighter sources, which is not the case at the fainter limit.

 We also noticed that the spatial distribution of new sources detected in this work is similar to the sources having counterparts in L20 or D23. This result suggests that the new sources are distributed throughout the field.

%{\st {We also conducted a comparison of the spatial positions of the new sources in each of the fields. We computed the median RA and DEC from all the sources detected within each field. Subsequently, we calculated the fraction of the number of sources within annular regions centered at median RA and DEC. The distribution is constructed for new sources of this work and sources having counterparts in L20 or D23. The comparison histogram is provided in Figure. The resemblance between the distribution of new sources and those with counterparts in D23 or L20 strongly suggests that the new sources are evenly distributed throughout the field.}}

%In Figure~\ref{Additional_SRC_COMP}, we have provided the fraction of our extra source as a function of magnitude for both the comparison fields. It illustrates that the methodology we have currently employed is capable of detecting a substantial number of low-magnitude sources. This capability effectively extends the limits of source detection achievable using the AstroSat-UVIT, providing valuable insights into the fainter UV sources.

\begin{figure}
\centering
\includegraphics[angle=0,height=6cm,width=8cm]{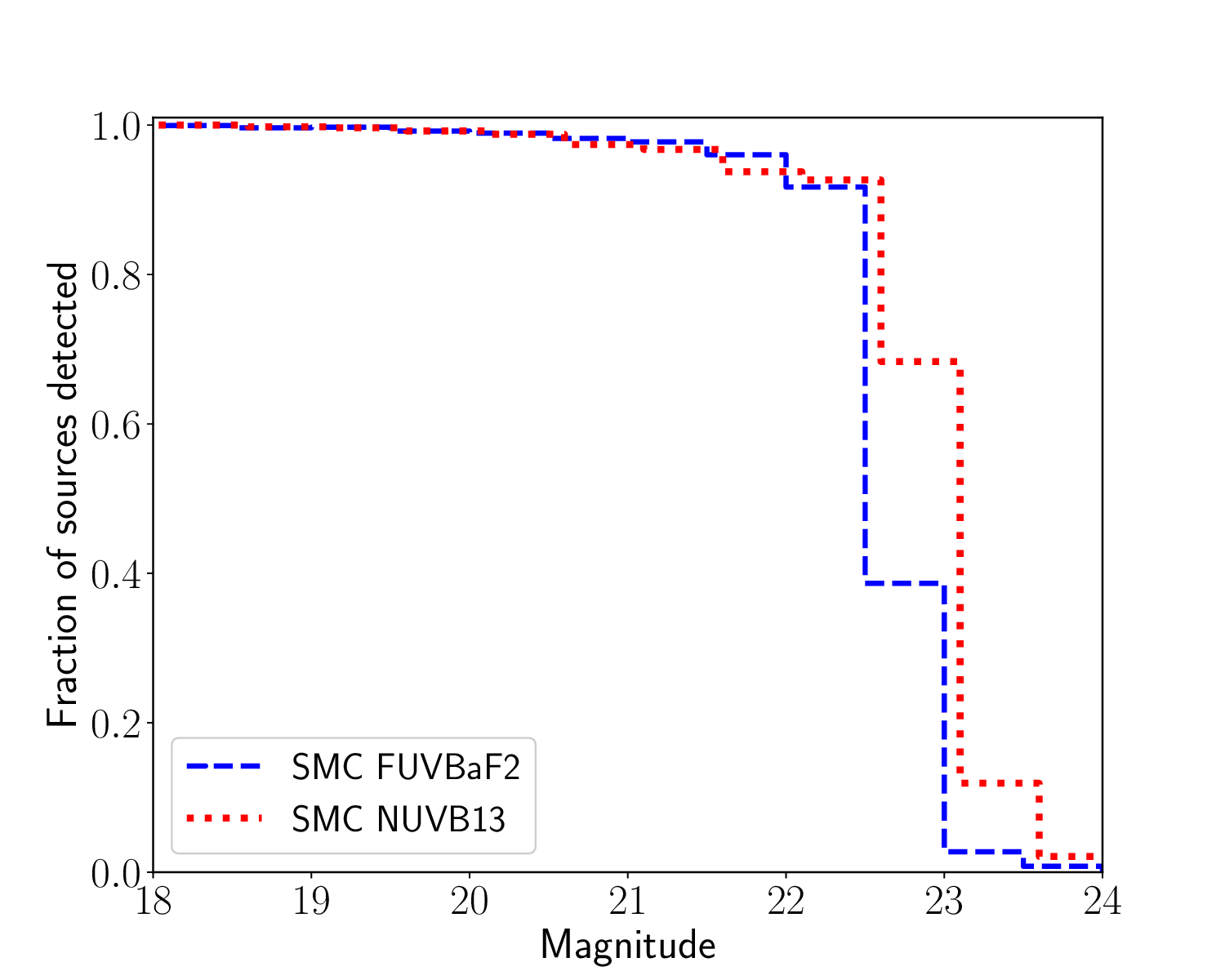}
\caption{Completeness of the catalog as a function of magnitude in the field of SMC. \label{Completeness}}

\end{figure}

\begin{figure*}
\centering
\includegraphics[angle=0,height=6cm,width=16cm]{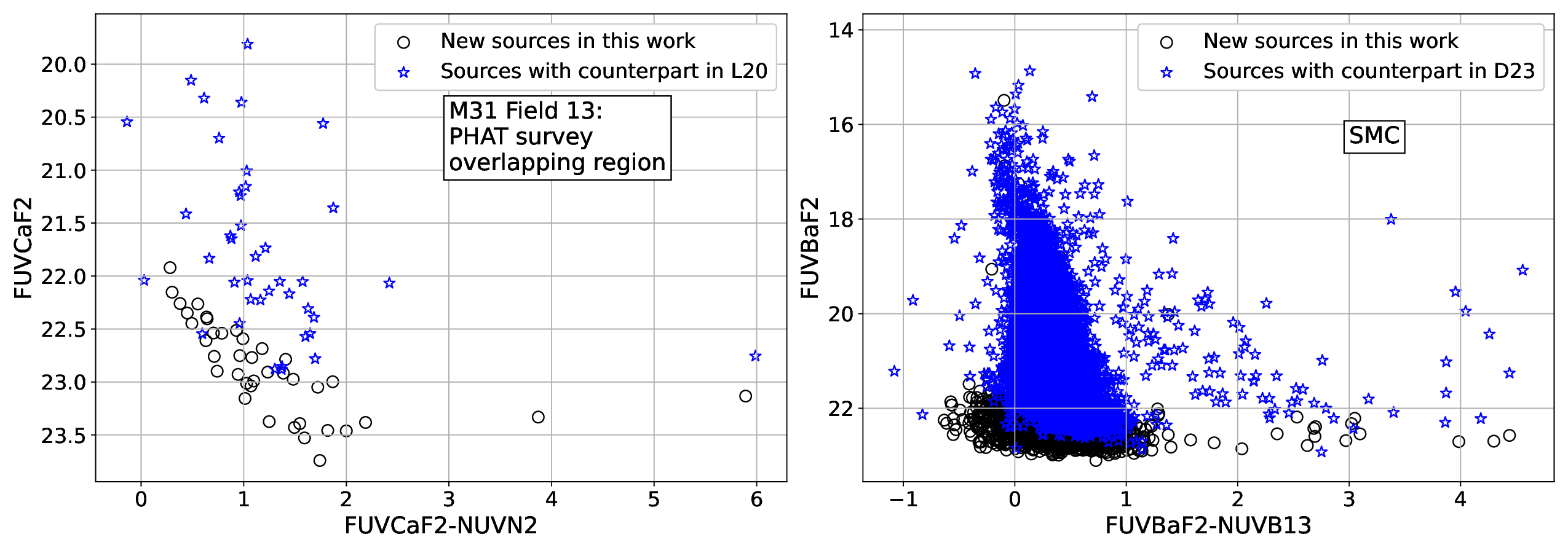}
\caption{ Color-magnitude diagram for sources detected in both FUV and NUV having counterparts in previous UVIT studies compared with new sources detected in both FUV and NUV of this work. Left panel: In M31 Field 13 compared with L20 \citep{Leahy2020}. Right panel: In SMC compared with D23 \citep{Ashish2023}.  \label{CMD}}

\end{figure*}

\subsection{Completeness}
Completeness is derived by recovering the artificial stars added in each field of SMC \citep[e. g.,][]{Ashish2023}. For each field, the psf is constructed using the {\it psf} task in Image Reduction and Analysis Facility {\it IRAF}\footnote {https://iraf.net/}. We used the {\it addstar} routine of {\it IRAF} to introduce synthetic sources consistent with this psf. For each trial, we constrained the number of stars added to a fraction not exceeding 10$\%$ of the total sources. These sources are introduced within each magnitude bin, and we employed our current methodology to detect them post-addition. The fraction of sources recovered as a function of magnitude is provided in Figure~\ref{Completeness}. The catalog is $\sim 92$-$93\%$ complete between $22$-$22.5$ mag in FUVBaF2 and NUVB13 filters.

%Simple mathematics can be inserted into the flow of the text e.g. $2\times3=6$
%or $v=220$\,km\,s$^{-1}$, but more complicated expressions should be entered
%as a numbered equation:

%\begin{equation}
%    x=\frac{-b\pm\sqrt{b^2-4ac}}{2a}.
%	\label{eq:quadratic}
%\end{equation}

%Refer back to them as e.g. equation~(\ref{eq:quadratic}).

%\begin{table}
%	\centering
%	\caption{This is an example table. Captions appear above each table.
%	Remember to define the quantities, symbols and units used.}
%	\label{tab:example_table}
%	\begin{tabular}{lccr} % four columns, alignment for each
%		\hline
%		A & B & C & D\\
%		\hline
%		1 & 2 & 3 & 4\\
%		2 & 4 & 6 & 8\\
%		3 & 5 & 7 & 9\\
%		\hline
%	\end{tabular}
%\end{table}
\subsection{Color-magnitude diagram}
A comparative study of the color-magnitude diagram (CMD) for sources detected in both FUV and NUV bands in this work and that of L20 and D23 catalogs was carried out. It is noticed (Figure~\ref{CMD}) that the new sources found in this work follow a similar trend as that observed in the case of common sources.

\subsection{Crowded region}
It is observed that in the crowded region, the detection efficiency is low compared to the L20. In the overcrowded regions of M31, we detected fewer sources than L20. The reason could be our conservative approach in using the deblending parameter to separate the nearby sources in crowded regions and the CLEAN parameter in SExtractor, which reduces spurious detections near bright sources. We have not attempted to analyze the crowded region in this work.

\begin{figure*}
\centering
\includegraphics[angle=0,height=9cm,width=16cm]{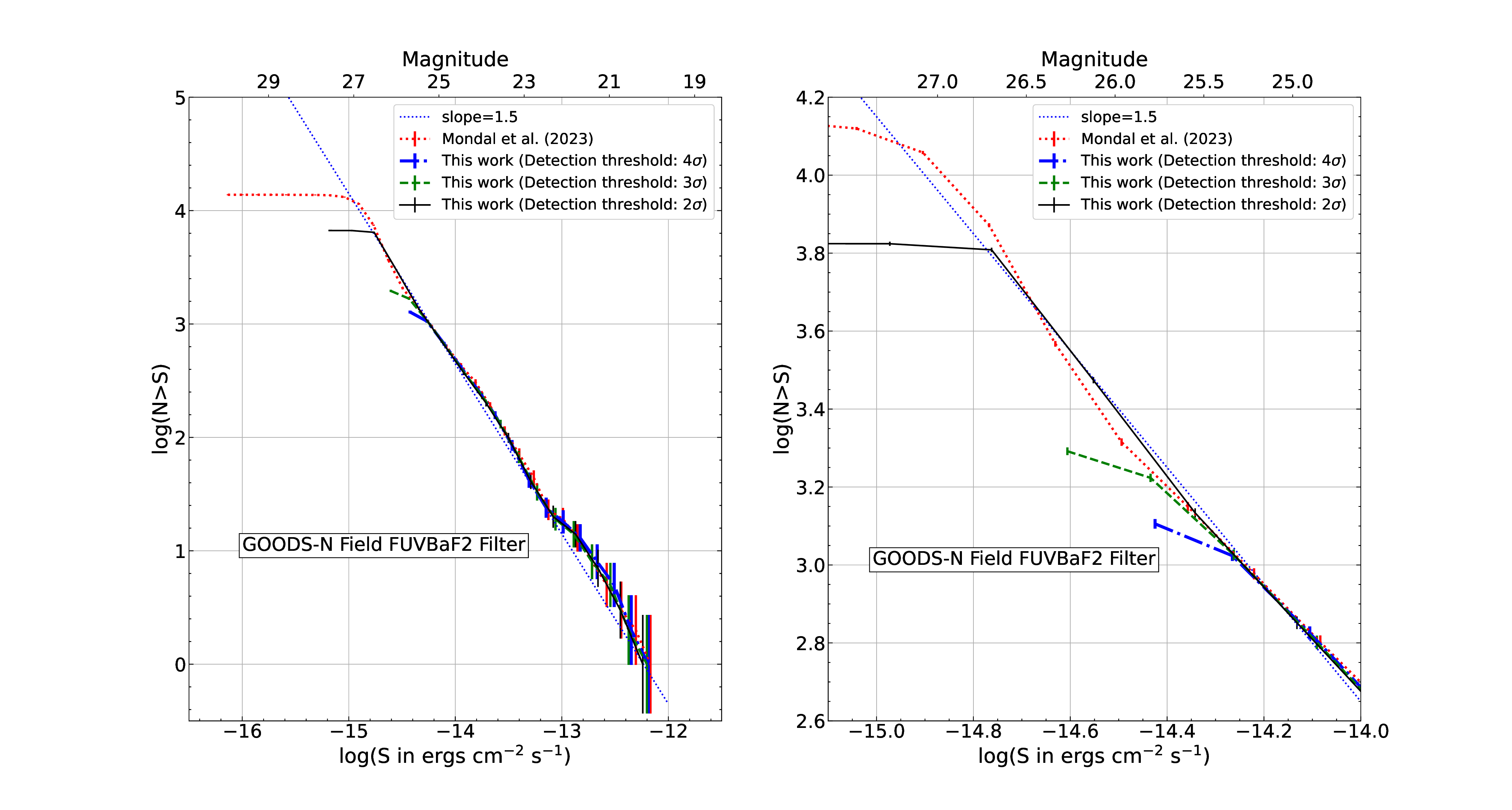}
%\gridline{\fig{LogNvs_logs_Mag_Our_src_GOODS.eps}{0.45\textwidth}{}
%          \fig{LogNvs_logs_Mag_Our_src_GOODS_zoom.eps}{0.45\textwidth}{}}
 \caption{Left panel: Log(N) vs. Log(S) plot in the FUVBaF2 filter of the GOODS-N field for detected sources in this work at $4$, $3$ and $2\sigma$ equivalent threshold and sources detected in \citet{Chayan2023}. A straight line of slope $1.5$ is plotted as a reference. Right panel: Zoomed-in image of Figure~\ref{lognlogsgoods} showing the magnitude limits reached at different thresholds. \label{lognlogsgoods}}
   
\end{figure*}

\subsection{Comparison in GOODS-N field:} \label{GOODS_NORTH_Mondal}

We also applied our method in deep UVIT observations in the FUVBaF2 filter of the Great Observatories Origins Survey Northern (GOODS-N) field. We compared our results with the catalog of \citet{Chayan2023} utilizing the same observation. With an exposure time of $\sim 34$ kiloseconds, in the FUVBaF2 filter, they reached a $50\%$ completeness limit of $26.4$ magnitude. We observed that on reducing the threshold to $2\sigma$ equivalent or below, the source distribution of our catalog matches well with \citet{Chayan2023}. The log(N) vs. log(S) plot for our detected sources at $ 2 \sigma$, $3 \sigma$, and $4 \sigma$ equivalent thresholds, and with the catalog of \citet{Chayan2023} are provided on the left panel of Figure~\ref{lognlogsgoods}. The zoomed-in figure of the same near the magnitude limits for various thresholds is shown on the right panel of Figure~\ref{lognlogsgoods}. For this deep exposure, the slope of the log(N) vs. log(S) plot is consistent with a slope of $1.5$. This result indicates that faint sources from the catalog of \citet{Chayan2023} have $2\sigma$ or lower detection significance.

%{\bf We have also generated a color-magnitude diagram (CMD) for the sources (Figure~\ref{CMD}) that have been detected in both NUV and FUV. We conducted a comparison between the CMD obtained from the L20 and D23 catalogs and additional sources obtained from our analysis. A detailed study of the properties of these sources is required to understand the true nature of these sources, which will be part of future work.}

%Apparently, in the case of M31, these additional sources exhibit a more NUV emission compared to what is observed in the SMC. This observed trend could be just due to the sensitivity limits of the filters, as the number of sources detected in NUV is more in the case of SMC and conversely in FUV for M31. Therefore, 

%\vspace{1cm}
\subsection{Summary \& Conclusions}
Given the low UV sky background, it is more appropriate to model the UVIT background with a Poisson distribution. 

Therefore, we considered the Poisson distribution for the background and threshold calculations and carried out the source detection. We restrict the catalog in this work to a 4$\sigma$ equivalent threshold for higher reliability. This catalog includes a substantial fraction of sources from previous studies (L20 and D23). However, we also detect a large number of new sources. Most of these new sources have counterparts in extensive catalogs at other wavebands, lending greater confidence in these detections. New sources detected are at the fainter magnitude limit, thereby pushing the limits for detecting the sources from UVIT. This study has shown that a more appropriate treatment of the background can yield an increase in the source number by a factor of $\sim 1.5-2$. This approach can form a basis for a more complete and comprehensive UVIT catalog. Further studies on the nature of these new faint sources will presented elsewhere.

\vspace{1cm}

\begin{acknowledgments}
BA, DB, and SB thank the Department of Space, Govt of India, for the financial support under the ISRO RESPOND Project (grant number No.DS$\_$2B-13012(2)/43\\/2017).
BA also acknowledges the DST-INSPIRE program (Reg. No.: IF190146) for the funding under which some of this work has been carried out. The authors thank Prof. C. S. Stalin, IIA, Bangalore for the fruitful discussion.
The research is based to a significant extent on the results
obtained from the AstroSat mission of the Indian Space Research
Organization (ISRO), archived at the Indian Space Science Data Centre
(ISSDC). The Payload Operations Centre at IIA processed the UVIT data used here. The UVIT is
built in collaboration between IIA, IUCAA, TIFR, ISRO and CSA. This work has made use of data from the European Space Agency (ESA) mission
{\it Gaia} (\url{https://www.cosmos.esa.int/gaia}), processed by the {\it Gaia}
Data Processing and Analysis Consortium (DPAC,
\url{https://www.cosmos.esa.int/web/gaia/dpac/consortium}). Funding for the DPAC
has been provided by national institutions, in particular the institutions
participating in the {\it Gaia} Multilateral Agreement. Manipal Centre for Natural Sciences, Centre of Excellence, Manipal Academy of Higher Education (MAHE) is acknowledged for facilities and support.
\end{acknowledgments}
%% in the research for the paper. Each keyword is check against the master 
%% list during copy editing. Individual instruments can be provided in 

\vspace{5mm}
\facilities{{AstroSat, HST(WFC3), GALEX, Gaia}}


\software{
Source Extractor \citep{Bertin1996}, SCAMP \citep{Bertin2006}, CCDLAB \citep{Postma2017, Postma2021}, SAOImageDS9 \citep{ds92003}, TOPCAT \citep{Taylor2005}, Matplotlib \citep{Matplotlib2007}, Astropy \citep{Astropy2013,Astropy2018}.}\\



\bibliography{astro_ref}{}
\bibliographystyle{aasjournal}

%% the collaboration and author truncation commands are used. It has to


\end{document}